\RequirePackage{fix-cm}
\documentclass[smallextended]{svjour3}     
\smartqed 
\usepackage{footnote}
\usepackage{tablefootnote}
\usepackage{float}
\usepackage{graphicx}
\usepackage{amsmath}
\usepackage{cite}
\usepackage{multirow}
\usepackage{braket}
\usepackage{amsfonts}
\usepackage{bbold}
\begin{document}
\title{Contextuality based  quantum conferencing}
\author{Rajni Bala     \and Sooryansh Asthana  \and   V. Ravishankar
}
\institute{Rajni Bala \at Department of Physics, Indian Institute of Technology Delhi, New Delhi-110016, India, \\\email{Rajni.Bala@physics.iitd.ac.in} \and Sooryansh Asthana  \at Department of Physics, Indian Institute of Technology Delhi, New Delhi-110016, India, \\\email{sooryansh.asthana@physics.iitd.ac.in} \and  V. Ravishankar
 \at Department of Physics, Indian Institute of Technology Delhi, New Delhi-110016, India, \\ \email{vravi@physics.iitd.ac.in}
}

\date{Received: date / Accepted: date}

\maketitle

\begin{abstract}
Nonlocality inequalities for multi-party systems act as contextuality inequalities for single qudit systems of suitable dimensions [Heywood and Redhead, Found. Phys., 13(5), 481–499, 1983; Abramsky and Brandenburger,  New J. Phys., 13(11), 113036, 2011]. In this paper, we propose the procedure for adaptation of nonlocality-based quantum conferencing protocols (NQCPs) to contextuality-based QCPs (CQCPs). Unlike the NQCPs, the CQCPs do not involve nonlocal states.   As an illustration of the procedure,  we present a QCP based on Mermin's contextuality inequality. As a significant improvement, we propose a QCP based on CHSH contextuality inequality involving only four-dimensional states irrespective of the number of parties sharing the key. The key generation rate of the latter is twice that of the former. Although CQCPs allow for an eavesdropping attack which has no analog in NQCPs, a way out of this attack is demonstrated. Finally, we examine the feasibility of experimental implementation of these protocols with orbital angular momentum (OAM) states.
\end{abstract}

 \section{Introduction} 
\label{Introduction}
The surge of interest in nonclassical aspects of quantum mechanics owes largely to the applications that they offer. Examples include search algorithms and quantum computing algorithms \cite{Grover,deutschjozsa} which provide computational speedups as compared to their classical counterparts. There also exist altogether novel applications such as  quantum teleportation, superdense coding \cite{Bennett92a, Bennett93}, and quantum key distribution protocols. In particular, fundamental features of quantum mechanics were first applied in quantum key distribution (QKD)  \cite{Bennett84, Ekert91,namkung2020generalized}. Thanks to the no-cloning theorem \cite{Wootters82} and the existence of non-orthogonal bases, QKD protocols are unconditionally secure. In the celebrated Ekert protocol \cite{Ekert91}, nonlocal states are employed for key distribution. Though these states are not essential in the generation and certification of the key, they play a crucial role in device-independent QKD protocols \cite{Vazirani14}. Since nonlocality is a costly resource, other QKD protocols have been proposed whose security analyses are based on monogamy relations of other quantum features, e.g., quantum discord \cite{su2014Guassiandiscord,pirandola2014quantum} and contextuality \cite{Troupe15, Singh17}. Additionally, the Kochen-Specker theorem has been identified as a condition for secure QKD in \cite{Nagata05}. The contextuality-based QKD protocols proposed in \cite{Troupe15, Singh17} can be employed to generate a  secure key between two parties.

It has been noticed \cite{Heywood83, Mermin90a, Guhne10} that the CHSH nonlocality inequality for a bipartite system detects contextuality in a single system. The interrelation between nonlocality and contextuality has been further placed on a firm ground using the sheaf-theoretic framework \cite {Abramsky11}. In short, nonlocality inequalities can be adapted to detect contextuality in a single system of suitable dimensions. The reason underlying this adaption is the isomorphism of Hilbert spaces of identical dimensions. 

\begin{figure}[ht!]
    \centering
    \includegraphics[width=0.7\textwidth]{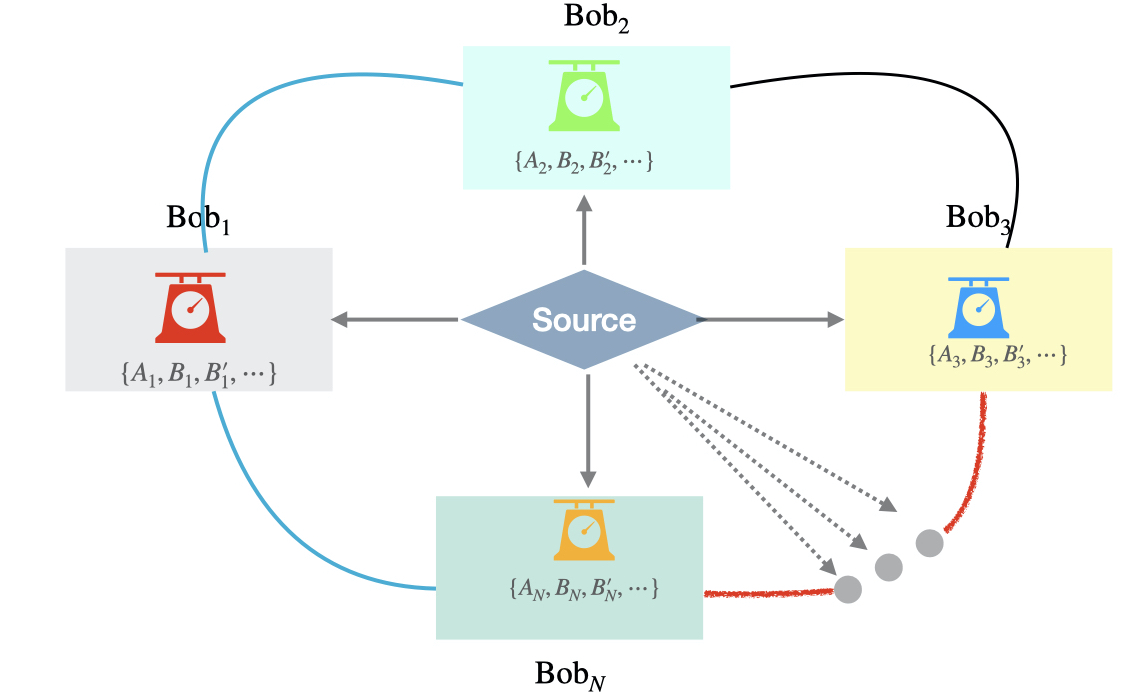}
    \caption{Pictorial representation of QCP based on nonlocality inequality: an $N$-party nonlocal  state is shared among $N$ parties, {\it viz.}, ${\rm Bob}_1, \cdots, {\rm Bob}_N$ who perform random measurements of  $A_k, B_k,B_k'\cdots ; 1\leq k \leq N$. The outcomes of $ B_k,B_k'\cdots$ are revealed on a classical channel to check for violation of nonlocality inequality and those of $A_k$ are used to generate the key.}
    \label{fig:entanglement}
\end{figure}

\begin{figure}[ht!]
    \centering
    \includegraphics[width=1.0\textwidth]{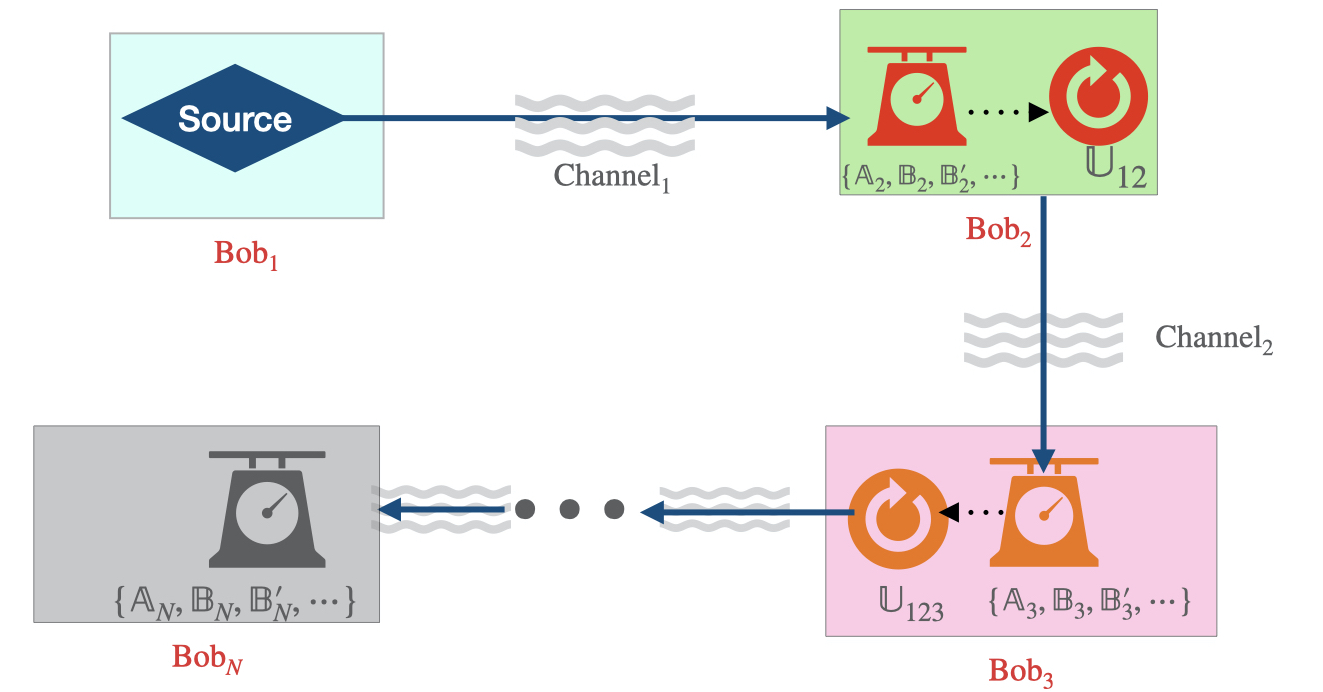}
    \caption{Pictorial representation of QCP based on contextuality: ${\rm Bob}_1$ has a source to produce a $ d^N$-- dimensional state.  Bob$_k$ performs random measurement of one of the observables from the set $\{\mathbb{A}_k, \mathbb{B}_k, \mathbb{B}_k',\cdots\}$ and thereafter, performs an arbitrary unitary transformation $\mathbb{U}_{1\cdots k}$ to make the protocol resilient against such attacks in which Eve's observables commute with those of subsequent Bobs. The transformed state is then sent to Bob$_{k+1}$. The wavy lines in the figure represent possible noise in the channel.}
    \label{fig:contextuality}
\end{figure}

In this paper,  we exploit the formal equivalence between nonlocality and contextuality to propose QCPs based entirely on contextuality, corresponding to any NQCP. However, our protocols are significantly different from the NQCPs. The isomorphism between the Hilbert spaces extends only upto the states and the algebra of the observables. The implementation of the QCPs and security against eavesdropping are completely different. In short, the CQCPs are not just mathematical analogs of NQCPs. So, the similarity and contrast of the two QCPs are schematically shown in figures (\ref{fig:entanglement}) and (\ref{fig:contextuality}) respectively.  Particular attention may be paid to ``masking transformations'' $\mathbb{U}$, which are employed for security, in figure (\ref{fig:contextuality}). It is discussed in detail in section (\ref{Security}). 

 We propose two classes of CQCPs.  In the first class, a higher dimensional state is sent from the first party to the last party sequentially. In the second class, the parties are divided into partially overlapping groups and a relatively lower dimensional state is used for sharing the key among different groups. To illustrate these two classes, we have explicitly described QCPs based on Mermin's contextuality inequality and CHSH contextuality inequality respectively.

 Since all these protocols involve higher dimensional states, orbital angular momentum (OAM) states of light  seem to be natural candidates for their implementation. There have been numerous recent advances in the generation \cite{beijersbergen94, Heckenberg92} and manipulation of higher dimensional OAM states of light, which provide an edge to quantum information processing with  higher dimensional states \cite{Mair01, leach2002measuring, berkhout2010efficient, Dada11, Fickler12, malik2014direct, zhou2016orbital, yin2017, erhard18}. Noting this, we briefly outline how CQCPs may possibly be realised experimentally using OAM states.

The paper is organised as follows: in section (\ref{Notation}), we set up the notation to be used in the paper, for an uncluttered discussion. In section (\ref{NL_contextality}), to make the discussion easier, the  properties probed by CHSH contextuality inequality in a single system have been discussed. 
Section (\ref{Procedure}) develops the procedure, which is central to the paper, for obtaining CQCP from any NQCP. Section (\ref{Security}) discusses  eavesdropping strategies and possible ways out. In section (\ref{Protocol}), we apply the procedure to propose the QCP based on Mermin's contextuality inequality and discuss its key generation rate. Section (\ref{Location}) discusses how one can pinpoint the location of Eve.  In section (\ref{Bell_Protocol}), the QCP based on CHSH contextuality inequality and its key generation rate have been discussed.   Section (\ref{OAM}) presents a possibility of how  the protocols may be implemented using orbital angular momentum (OAM) states of light. In section (\ref{Error}), effects of imperfect preparation of the states and noisy measurements on the QCP based on Mermin's and CHSH contextuality inequality are studied. Section (\ref{Conclusion}) summarises the paper with concluding remarks. 

 \section{Notation}
 \label{Notation}
 In this section, we set up the notation to be used henceforth in the paper:
 \begin{enumerate}
 
     \item\begin{enumerate}
         \item Observables acting on a multi-party system will be represented by $A_k, A'_k, \cdots$. The party which measures these observables is represented in the subscript.
         \item  The corresponding observables acting on a single qudit system will be represented by $\mathbb{A}_k, \mathbb{A}_k', \cdots$.
     \end{enumerate}
     \item {\bf Mapping between bases:}\\ Let 
\begin{eqnarray}
&\mathfrak{B}_1 &\equiv  \{|j_1j_2\cdots j_{N}\rangle; j_k\in \{0, 1,\cdots,d-1\}, 1\leq k \leq N\}; \nonumber\\
& \mathfrak{B}_2 &\equiv \{|0\rangle, |1\rangle, \cdots, |D-1\rangle\},
\end{eqnarray}     
 where $\mathfrak{B}_1$ is a basis for the tensor product space of $N$ $d$-dimensional Hilbert spaces $({\cal H}^d)^{\otimes N}$. Simialarly, $\mathfrak{B}_2$ is a basis for a Hilbert space, ${\cal H}^D$, of dimension $D\equiv d^N$. Since $({\cal H}^d)^{\otimes N}$ and ${\cal H}^D$  are isomorphic to each other, we set up the following bijective mapping between the bases $\mathfrak{B}_1$ and $\mathfrak{B}_2$:
     \begin{eqnarray}\label{eq:1}
         |j_1\cdots j_{N}\rangle \leftrightarrow \Big|\sum_{k=1}^{N}d^{N-k}j_{k}\Big\rangle=\ket{j}.
     \end{eqnarray}
     \item {\bf Mapping of observables:}\\ The symbol $\mathbb{M}_k$ represents the observable having the  same matrix representation in the  basis $\mathfrak{B}_2$, as the observable $\mathbb{1}^{\otimes k-1}\otimes M_k\otimes \mathbb{1}^{\otimes N-k}$ has in the basis $\mathfrak{B}_1$. That is,
     \begin{eqnarray}
     \label{Basischange}
         \langle j|\mathbb{M}_k|l\rangle &= \big\langle j_1\cdots j_N\big|\mathbb{1}^{\otimes k-1}\otimes M_k\otimes \mathbb{1}^{\otimes N-k}\big\vert l_1\cdots l_N\big\rangle.
     \end{eqnarray}
     In particular, the projection operator corresponding to an eigenvalue $m$ of an operator $\mathbb{M}$ will be represented by ${\mathbb{\Pi}}_{\mathbb{M}}^m$.
      \item We  denote the states of multi-party systems by lowercase Greek letters $|\psi\rangle, |\xi\rangle, \cdots$ and the states of a single qudit by uppercase Greek letters $|\Psi\rangle, |\Xi\rangle, \cdots$.
      
      \item \textbf{Mapping of unitary transformations:}\\ \begin{enumerate}\item 
      Employing the mapping defined in equation (\ref{eq:1}), the unitary transformations performed by the $k^{th}$ party in a multi-party system, represented by $U_k$, are mapped to $\mathbb{U}_k$ in a single qudit, i.e.,
      \begin{eqnarray}
          \mathbb{1}^{\otimes k-1} \otimes U_k\otimes\mathbb{1}^{\otimes N-k}\leftrightarrow \mathbb{U}_k.
      \end{eqnarray}
      \item Similarly, the unitary transformations performed on the combined space of $j^{th}$ and $k^{th}$ party, represented by $U_{jk}$, are mapped to $\mathbb{U}_{j k}$ in a single qudit system, i.e.,
      \begin{eqnarray}
          \mathbb{1}^{\otimes j-1}\otimes U_{jk}\otimes \mathbb{1}^{N-k}\leftrightarrow \mathbb{U}_{jk}.
      \end{eqnarray}
      
      Similarly, the transformation acting on the combined space of first $k$ parties of a multiparty system, $U_{1\cdots k}$, is mapped to  $\mathbb{U}_{1\cdots k}$ in a single qudit system straightforwardly, i.e.,
      \begin{eqnarray}
          U_{1\cdots k}\otimes {\mathbb{1}}^{N-k}\leftrightarrow \mathbb{U}_{1\cdots k}.
    \end{eqnarray}
         \end{enumerate}
          \item The symbols $X_k, Y_k, Z_k$ shall be reserved for the Pauli matrices acting over the space of the $k^{\rm th}$ qubit and their counterparts for a single qudit will be represented by $\mathbb{X}_k, \mathbb{Y}_k, \mathbb{Z}_k$.
     \item Finally, we represent the sequential measurements of observables $A_1, \cdots, A_N$, such that $A_1$ is measured first and $A_N$ is measured last, by the symbol $(A_1\cdots A_N)$.
 \end{enumerate}
 

            \section{Relation between multi-party nonlocality and contextuality in a single qudit}
          \label{NL_contextality}
           Multi-party states that violate nonlocality inequalities also exhibit contextual behaviour, i.e., nonlocality and contextuality coexist in the multi-party states \cite{Guhne10}. However, multi-party nonlocality inequalities, with the mapping given in equation  (\ref{eq:1}), can be adapted to detect contextuality in a single qudit system of suitable dimensions. In this section, for the purpose of pedagogy, we illustrate this with the example of CHSH inequality.

          \subsection{CHSH inequality as a contextuality inequality}
          \label{Bell_contextuality} 
           CHSH inequality detects nonlocality in a bipartite system of arbitrary dimension \cite{Clauser69}. We show that the same inequality also detects contextuality. Consider a simple example of two-qubit systems. Let $\ket{\psi}$ be the singlet state of two-qubits, i.e., $|\psi\rangle=\frac{1}{\sqrt{2}}(|01\rangle-|10\rangle)$. Consider a set of three observables $\{X_1, X_2, Z_2\}$. 
 Evidently, $[X_1, X_2] = [X_1, Z_2] =0$ but $[X_2, Z_2] \neq 0$. We are interested in the following two contexts:\\
 \textbf{Context 1:} $(X_2X_1)$, i.e., $X_2$ is measured first followed by  $X_1$. If the outcome of $X_2$ is $+1  (-1)$, the outcome of  $X_1$ is guaranteed to be $-1 (+1)$.\\
 \textbf{Context 2:} $(Z_2X_1)$. Irrespective of the outcome of  $Z_2$, which is $\pm1$, the outcome of $X_1$ can be $+1$ or $-1$ with equal probability.\\
 Thus, the outcome of $X_1$ depends on the observable with which it is measured, i.e., it depends on the {\it context}. Thus, CHSH nonlocality inequality also detects contextuality  in a two-qubit system.\\
 
 Of course, a single qudit cannot exhibit nonlocal behaviour. In what follows, we show how a similar contextual behaviour gets manifested  in a single four-level system. Employing the following mapping between the bases of a two-qubit system and a single four level system,
 \begin{eqnarray}
     |00\rangle\equiv|0\rangle;~|01\rangle\equiv |1\rangle;~|10\rangle\equiv|2\rangle;~|11\rangle\equiv|3\rangle,
 \end{eqnarray}
 the singlet state $|\psi\rangle$ gets mapped to $|\Psi\rangle$,
 \begin{eqnarray}
 |\Psi\rangle= \frac{1}{\sqrt{2}}(|1\rangle-|2\rangle),
 \end{eqnarray}
 and the observables $X_1, X_2, Z_2$ map to the following observables:
      \begin{eqnarray}
         X_1\mapsto &\mathbb{X}_1= |0\rangle\langle 2|+|2\rangle\langle 0|+|1\rangle\langle 3|+|3\rangle\langle 1|\nonumber\\
        X_2\mapsto  &\mathbb{X}_2= |0\rangle\langle 1|+|1\rangle\langle 0|+|2\rangle\langle 3|+|3\rangle\langle 2|\nonumber\\
        Z_2 \mapsto &\mathbb{Z}_2=|0\rangle\langle 0|-|1\rangle\langle 1|+|2\rangle\langle 2|-|3\rangle\langle 3|.
      \end{eqnarray}
Naturally, the isomorphism of Hilbert spaces extends to the algebra of operators.

We may consider the same contexts  for $\mathbb{X}_1, \mathbb{X}_2, \mathbb{Z}_2$ with the state $|\Psi\rangle$ as considered for $X_1, X_2, Z_2$ with the singlet state $|\psi\rangle$. \\
       \textbf{Context 1 ($\mathbb{X}_2\mathbb{X}_1$):} i.e., $\mathbb{X}_2$ is measured first followed by measurement of $\mathbb{X}_1$. If the outcome of $\mathbb{X}_2$ is $+1 (-1)$,  the outcome of $\mathbb{X}_1$ is guaranteed to be $-1 (+1)$.\\
           \textbf{Context 2 ($\mathbb{Z}_2\mathbb{X}_1$):}  Irrespective of the outcome of $\mathbb{Z}_2$, the outcome of $\mathbb{X}_1$ will be $+1$ or $-1$ with equal probability.\\
              The outcome of $\mathbb{X}_1$ depends on which observable has been measured before it, i.e., it depends on the {\it context}. So, the isomorphism extends to the contexts as well.  Thus, CHSH inequality can be adapted to detect sequential contextuality in a single four-level system for which it can be referred to as CHSH contextuality inequality. Similarly, $N$--party Mermin's nonlocality inequality can be referred to as Mermin's contextuality inequality in a single qudit system of appropriate dimension \cite{Abramsky11}, as illustrated in appendix (\ref{Mermin_contextuality}). 
  \section{The procedure for obtaining contextuality-based QCP from any nonlocality-based QCP}
  \label{Procedure}
  
 In this section, we employ the formal equivalence between contextuality and nonlocality to develop the procedure to obtain a CQCP from a NQCP. 
 
  Let $\{A_k,B_k, B'_k, B''_k, \cdots\}$ be a set of observables, acting over the $k^{\rm th}$ qudit of an $N$-qudit system. Any nonlocality inequality can be cast into the form \cite{Brunner14}, 
\begin{eqnarray}\label{eq:o}
\centering
\langle O\rangle \leq c,
\end{eqnarray} 
       where $O$ consists of sums of multilinears in the observables $B_k, B'_k, B''_k$ and $c$ is a non-negative number. Consider a NQCP in which  the violation of the inequality (\ref{eq:o}) acts as a security check and the outcomes of the observables $\{A_1, \cdots, A_N\}$ are used to generate the shared secret key. The sets $\{A_1, \cdots, A_N\}$ and $\{B_1, B'_1, \cdots, B_N, B'_N\}$ may be partially overlapping or completely disjoint, examples being Ekert protocol \cite{Ekert91} and Mermin inequality based QKD protocol \cite{MABK} respectively.
\begin{table}[ht!]
   \begin{tabular}{|c|c|c|} 
 \hline
{\bf S. No.}&{\bf NQCP} & {\bf CQCP}\\
  [0.5ex] 
 \hline\hline
 1 &  All the observables are  & All the observables are\\
 & publicly announced. & publicly announced.\\
 \hline
2 & $|\psi\rangle:$ $N$--qudit nonlocal state & $|\Psi\rangle:$ a single $d^N$-dimensional\\  & simultaneously shared among $N$   &  reference state. \\
& parties {\it viz.}, Bob$_1,\cdots,\rm{Bob}_N$.&\\
\hline
\multirow{23}{0.45em}{3} & \multirow{24}{8em}{ Bob$_k$  randomly  measures  one of the observables from the set $\{A_k, B_k, B'_k, \cdots\}$. }  &{\bf a)}~Bob$_1$ randomly prepares a state \\
&  &  $|\Psi'\rangle= \mathbb{U}_1{\mathbb{\Pi}}_{\mathbb{M}}^{{m}}|\Psi\rangle$, with a probability $\frac{1}{t}\langle \Psi|{\mathbb\Pi}_{\mathbb{M}}^m|\Psi \rangle $,    \\
& & where $\mathbb{M}$ is one of the observables from the set, \\
& &  $ \{\mathbb{A}_1, \mathbb{B}_1, \mathbb{B}'_1,\cdots\}$,  whose cardinality is $t$. The \\
& &  symbol $\mathbb{U}_1=e^{i(\alpha_1\mathbb{A}_1+\beta_1\mathbb{B}_1+\cdots)}$ represents an \\
& &   arbitrary unitary transformation, which commutes  \\
& &  with observables of subsequent Bobs.\\
  \cline{3-3}
&  & {\bf b)}~~ The state $|\Psi'\rangle$ is sent to\\
 & & Bob$_2$ who randomly measures one  \\
 & &  of the observables $\{\mathbb{A}_2, \mathbb{B}_2, \mathbb{B}'_2,\cdots\}$. \\
   \cline{3-3}
  \label{T2}
 & & {\bf c)} ~~He performs an arbitrary transformation\\
& &  $\mathbb{U}_{12}$ (which commutes with the observables\\
& &  of subsequent Bobs) on the post-measurement \\
& &  state and sends the transformed state to Bob$_3$.\\
  \cline{3-3}
&  & {\bf d)} This process continues until Bob$_N$\\
 & &  randomly measures one of the\\
 & &observables $\{\mathbb{A}_N, \mathbb{B}_N, \mathbb{B}'_N,\cdots\}$.\\
  \hline
  4 &This process is repeated& This process is repeated\\
   & for many  rounds.&for many rounds.\\
   \hline
5 &The violation of nonlocality & The violation of contextuality\\
 &  inequality $\langle{O}\rangle \leq c \implies$  &  inequality $\langle\mathbb{ O}\rangle \leq c \implies$ \\
 &  No eavesdropping (and vice versa). & No eavesdropping (and vice versa).\\
   \hline
6 & If the inequality is violated,& If the inequality is violated,\\
& the outcomes of $A_1,\cdots,A_N$ & the outcomes of $\mathbb{A}_1,\cdots,\mathbb{A}_N$  \\

  & act as a shared key. & act as a shared key.\\
 [1ex] 
 \hline
\end{tabular}
  \caption{Procedure for obtaining a CQCP from NQCP.}
    \label{Table_Procedure}
    
\end{table}
Let the corresponding contextuality inequality and the sets of observables be,
 \begin{eqnarray}
     \langle \mathbb{O} \rangle\leq c,
 \end{eqnarray} 
 \noindent and $\{\mathbb{A}_1, \cdots, \mathbb{A}_N\}$,  $\{\mathbb{B}_k,\mathbb{B}_k',\mathbb{B}_k^{''},\cdots\}$ respectively for a single qudit system of dimension $d^N$. That is to say, the observables $\mathbb{A}_k, \mathbb{B}_k$ and $A_k, B_k$ have the same representations in the bases $\{|0\rangle, \cdots, |d^N-1\rangle\}$ and $\{|j_1\cdots j_N\rangle; j_k \in \{0, 1, \cdots, d-1\}, 1\leq k \leq N\}$ respectively.
 
  The steps to obtain the corresponding CQCP are neatly summarised in table (\ref{Table_Procedure}),  and are further employed to propose the QCP based on Mermin's contextuality inequality in the section (\ref{Protocol}).
  
In the step 3(a) of table (\ref{Table_Procedure}), we have incorporated multiple operations into a single step. It is possible because a state is to be sent from Bob$_1$ to Bob$_2$ after measurement and necessary transformation. Bob$_1$ has no need to prepare a state $\ket{\Psi}$ first and then perform measurement and transformation over it. Instead, he can directly prepare a state $\ket{\Psi'}$ to be sent to Bob$_2$ which saves a lot of cost.

 \section{Security Analysis}  \label{Security}
 As we have seen in table (\ref{Table_Procedure}), some additional steps are performed in CQCPs.  These steps are needed to make the CQCPs resilient against eavesdropping. As the physical systems employed are completely different, the isomorphism between the two Hilbert spaces ({\it viz.}, ${\cal H}^{d^{\otimes N}} {\rm and}~ {\cal H}^{d^N}$) does not extend to eavesdropping strategies, and, hence, not to the security analyses. In this section, we discuss the security analyses of CQCPs for measurement-resend attacks by an eavesdropper. 
 
 Evidently, in the CQCPs, the measurement operators of all Bobs are mutually commuting. So, if Eve is present, say, between $k^{\rm th}$ and $(k+1)^{\rm th}$ Bob, her measurement operators may either commute or non-commute with those of the subsequent Bobs. In the latter case, any tampering by Eve gets reflected in the non-violation of contextuality inequality by the maximal amount. The former case, however, eludes a violation of contextuality inequality and Eve's tampering goes undetected. This case is unique to our protocols and merits a careful study in order to make our protocols robust against eavesdropping. 
  \label{Eavesdropping1}

 Assume that  Eve measures observables from the set $\{\mathbb{A}_\alpha,\mathbb{B}_\alpha,\mathbb{B}'_\alpha,\cdots;1\leq\alpha\leq k\}$, which naturally commute with those of $(k+1)^{\rm th}$ to $N^{\rm th}$ Bob.
  Measurements of these observables by Eve do not affect the outcomes of the observables of subsequent Bobs. Thus, there is no effect of these measurements on the violation of contextuality inequality. In this manner, Eve obtains full information about the key by choosing appropriate observables and still goes undetected. This point is further made more explicit with an example in appendix (\ref{Illustration}). 
  To obstruct Eve from obtaining any information, we propose a way out as explained in the following subsection.

  \subsection{The wayout: masking transformations}
  In order to make our protocol resilient against this attack, the following strategy may be adopted. After making a measurement on the state, each Bob performs a random unitary transformation   with the {\it sole condition} that it commutes with the observables of subsequent Bobs. That is to say, Bob$_1$ prepares a state and performs a random unitary transformation $\mathbb{U}_1$ generated by the observables accessed by him. Then, Bob$_2$ makes his measurement and performs a random unitary transformation  $\mathbb{U}_{12}$, generated by the observables accessed by both him and his predecessor. This chain is continued with every Bob but the last one. Thus, Bob$_k$ makes his measurement and performs a unitary transformation $\mathbb{U}_{1\cdots k}$ generated by the observables accessed by him and all his predecessors.  These random unitary transformations are termed as {\it masking transformations} as they mask the information about the outcomes of the observables of the preceding Bobs. Thus, Eve would be none the wiser about the key.

  We now make the procedure explicit for the protocol presented in table (\ref{Table_Procedure}). Bob$_1$, after preparing the state, say,  $\ket{\Psi_1}={\mathbb \Pi}_{\mathbb{A}_1}^{a_1}\ket{\Psi}$, performs a masking transformation,
  \begin{eqnarray}\label{eq:masking}
      \mathbb{U}_1= \exp{[{i(\beta_1\mathbb{B}_1+\beta'_1\mathbb{B}'_1+\cdots)}]};~ \beta_1,\beta_1'\in \mathbb{R}.
  \end{eqnarray}
  The transformed state, $\ket{\Psi'}=\mathbb{U}_1|\Psi_1\rangle$, is transmitted to Bob$_2$. Now, even if Eve intercepts the state on its way from Bob$_1$ to Bob$_2$ and performs a measurement of the observable $\mathbb{A}_1$, she does not gain information about the key. This is because the transformation $\mathbb{U}_1$ has changed the information about the outcome of observable $\mathbb{A}_1$, and thus, masks the information about the measurement of Bob$_1$. This feature holds at all the steps because each Bob (excluding the last one) performs a random unitary transformation. This justifies the nomenclature {\it masking} for these transformations.

  Masking transformations, together with the violation of contextuality inequality, assure the security of the protocol against such attacks.
  
  \subsection{Invariance of context under masking transformations}
  In what follows, we provide an explicit proof that masking transformations do not change the context and hence correlations in the outcomes of observables are intact.
  This is made possible by the fact that masking transformations performed by Bob$_k$ commute with the observables of subsequent Bobs (i.e., Bob$_{k+1},\cdots ,$ Bob$_{N})$ as guaranteed by the definition of $\mathbb{U}_1$ in equation (\ref{eq:masking}) (which commutes with observables of Bob$_2,\cdots ,$ Bob$_N$). An explicit proof follows.\\
\noindent{\it Proof:} The proof consists of three parts. In the first part, we show that the expectation values of observables of subsequent Bobs remain invariant after the masking transformation. In the second part, we show that the eigenstates of the observables of the subsequent Bobs lie within the same eigenspace even after the masking transformation.  In the last part, we show that correlations among different observables remain invariant.

\noindent{\bf (i) Invariance of expectation value of observables : }Let $\ket{\Phi_k}$ be the post--measurement state of Bob$_k$ which changes to $\ket{\Psi_k}$ after a masking transformation  $\mathbb{U}_{1\cdots k}$ has been performed by Bob$_k$, i.e.,
\begin{equation}\label{eq:unitary}
    |\Psi_k\rangle = \mathbb{U}_{1\cdots k}|\Phi_k\rangle.
\end{equation}
Note that, by definition, the masking transformation $\mathbb{U}_{1\cdots k}$ commutes with the observables of the subsequent Bobs, i.e., (Bob$_{k+1},\cdots,$ Bob$_N$). 
 As per the protocol, Bob$_{k+1}$ will perform a measurement of one of the observables from the set $\{\mathbb{A}_{k+1}, \mathbb{B}_{k+1}, \mathbb{B}'_{k+1}, \cdots\}$.  
We explicitly consider the observable $\mathbb{B}_{k+1}$ for which, 
\begin{align}
     \langle \Psi_{k}|\mathbb{B}_{k+1}|\Psi_k\rangle = \langle \Phi_k|\mathbb{U}^{\dagger}_{1\cdots k}\mathbb{B}_{k+1}\mathbb{U}_{1\cdots k}|\Phi_k\rangle = \langle \Phi_k|\mathbb{B}_{k+1}|\Phi_k\rangle.
\end{align}
Since the choice of $\mathbb{B}_{k+1}$ is arbitrary, this holds true for any observable of subsequent Bobs, including projection operators. Hence, all the probabilities and expectation values of the subsequent Bobs remain invariant under this transformation.

\noindent{\bf(ii) Invariance of outcomes of observables of subsequent Bobs:}  
In the protocol presented in the section (\ref{Procedure}), a key is generated when all the $N-$ Bobs perform measurements of their respective observables from the set $\{\mathbb{A}_1, \mathbb{A}_2, \cdots , \mathbb{A}_k, \mathbb{A}_{k+1}, \cdots, \mathbb{A}_N\}$. Given that, the post-measurement states of Bob$_k$ are eigenstates of the observables of the subset $\{\mathbb{A}_{k+1}, \cdots, \mathbb{A}_N\}$. We show that the states after the masking transformations continue to be the eigenstates of these observables with the same eigenvalues.

Suppose that the post-measurement state $|\Phi_k\rangle$ of Bob$_k$  is an eigenstate of $\mathbb{A}_{k+1}$ with eigenvalue $a_{k+1}$. Let $\ket{\Psi_k}$ be the state obtained after performing the masking transformation $\mathbb{U}_{1\cdots k}$ on $\ket{\Phi_k}$. We show that the state $|\Psi_k\rangle$ is also an eigenstate of $\mathbb{A}_{k+1}$ with the same eigenvalue $a_{k+1}$ as follows:
\begin{align}
    \mathbb{A}_{k+1}|\Psi_k\rangle = \mathbb{A}_{k+1}\mathbb{U}_{1\cdots k}|\Phi_k\rangle= \mathbb{U}_{1\cdots k}\mathbb{A}_{k+1}|\Phi_k\rangle
    =a_{k+1}\mathbb{U}_{1\cdots k}|\Phi_k\rangle= a_{k+1}|\Psi_k\rangle.\nonumber
\end{align}
The above equation holds due to the commutativity of $\mathbb{U}_{1\cdots k}$ with the observables of subsequent Bobs.
This clearly shows that if a state $\ket{\Phi_k}$ is an eigenstate of observable, so is the state $\ket{\Psi_k}$, which proves the assertion. Since the choice of the observable $\mathbb{A}_{k+1}$ is arbitrary, it implies that this result holds for any observable.

\noindent{\bf(iii) Invariance of correlations: } We prove this by demonstrating the invariance of correlation among three Bobs. It admits a straightforward generalization to $N-$ Bobs. Suppose that Bob$_1$ measures $\mathbb{\Pi}_{\mathbb{B}_1}^{b_1}$ on the state $\ket{\Psi}$, followed by a masking transformation $\mathbb{U}_1$ (generated by the observables accessible to Bob$_1$). Thereafter,  Bob$_2$ measures $\mathbb{\Pi}_{\mathbb{B}_2}^{b_2}$, followed by a masking transformation $\mathbb{U}_{12}$ (generated by the observables of Bob$_1$ and Bob$_2$). Finally, Bob$_3$ measures $\mathbb{\Pi}_{\mathbb{B}_3}^{b_3}$, followed by a masking transformation $\mathbb{U}_{123}$. The state, after this process, is given by,
\begin{align}
    \mathbb{U}_{123}{\mathbb{\Pi}}_{\mathbb{B}_3}^{b_3}\mathbb{U}_{12}{\mathbb{\Pi}}_{\mathbb{B}_2}^{b_2}\mathbb{U}_1{\mathbb{\Pi}}_{\mathbb{B}_1}^{b_1}|\Psi\rangle.
\end{align}
The correlation in the three outcomes is given by,
\begin{align}
      &  \langle \Psi|\mathbb{\Pi}_{\mathbb{B}_1}^{b_1}\mathbb{U}^{\dagger}_1\mathbb{\Pi}_{\mathbb{B}_2}^{b_2}\mathbb{U}^{\dagger}_{12}\mathbb{\Pi}_{\mathbb{B}_3}^{b_3}\mathbb{U}_{123}^{\dagger}\mathbb{U}_{123}{\mathbb{\Pi}}_{\mathbb{B}_3}^{b_3}\mathbb{U}_{12}{\mathbb{\Pi}}_{\mathbb{B}_2}^{b_2}\mathbb{U}_1{\mathbb{\Pi}}_{\mathbb{B}_1}^{b_1}|\Psi\rangle=\langle \Psi|\mathbb{\Pi}_{\mathbb{B}_1}^{b_1}\mathbb{\Pi}_{\mathbb{B}_2}^{b_2}\mathbb{\Pi}_{\mathbb {B}_3}^{b_3}|\Psi\rangle.
\end{align}
The above equation follows because of the commutativity among observables of different Bobs (in this case $\mathbb{B}_1,\mathbb{B}_2,\mathbb{B}_3$) and that of the masking transformations with those of subsequent Bobs, i.e., $[\mathbb{U}_1, \mathbb{\Pi}_{\mathbb{B}_j}^{b_j}] =0,~ j=2,3$;~$[\mathbb{U}_{12}, \mathbb{\Pi}_{\mathbb{B}_3}^{b_3}] =0$  and, $ [\mathbb{B}_i,\mathbb{B}_j]=0,~i,j=1,2,3$.
Thus, correlations remain intact after the application of a masking transformation.

The three results together show that masking transformations do not change the context. Thus, there is no effect of masking transformations either on  correlations in a key or on the violation of contextuality inequality.


\section{QCP based on Mermin's contextuality inequality} 
\label{Protocol}
\begin{figure}[ht]
    \centering
    \includegraphics[width=1.0\textwidth]{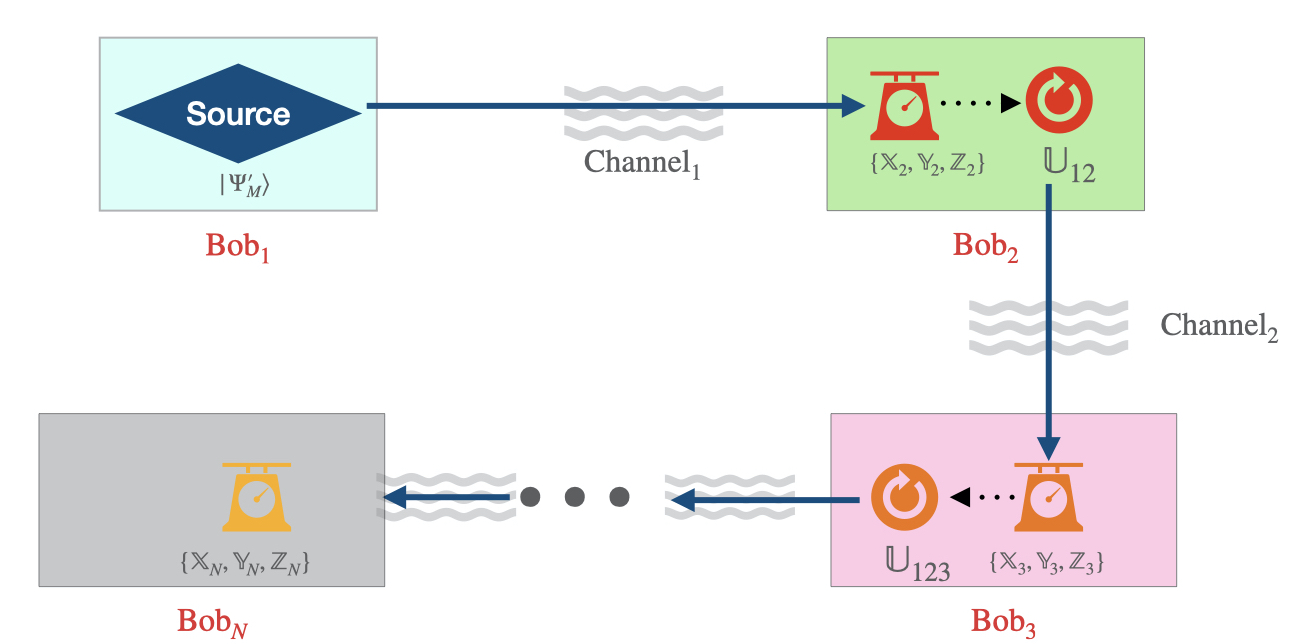}
    \caption{Schematic representation of Mermin's CQCP.
    }
    \label{fig:mesh1}
\end{figure}

 As the first illustration of the procedure mentioned in table (\ref{Table_Procedure}), we explicitly lay down the QCP based on Mermin's contextuality inequality \cite {Abramsky11} for distributing a key among $N$ parties, {\it viz.}, Bob$_1,\cdots,$
 $\rm{Bob}_N$. Each Bob can choose an observable from a set of three dichotomic observables, i.e., Bob$_k$ can choose from the set,
 \begin{eqnarray}
     {\cal S}_k\equiv \{\mathbb{X}_k,\mathbb{Y}_k,\mathbb{Z}_k\},~1\leq k\leq N.
 \end{eqnarray}
Given these observables, Mermin's contextuality inequality takes the form,
\begin{eqnarray}\label{eq:5}
    \mathbb{M}_N && =\frac{1}{2i}\Bigg|\bigg\langle\prod_{k=1}^{N} (\mathbb{X}_k+i \mathbb{Y}_k)-\prod_{k=1}^{N}(\mathbb{X}_k-i \mathbb{Y}_k)\bigg\rangle\Bigg|\leq c\nonumber\\
 c&&= 2^{N/2},\quad N={\rm even},\nonumber\\
 &&= 2^{\frac{N-1}{2}},\quad N= {\rm odd}.
\end{eqnarray}
Now, we are set to establish the protocol as per the procedure laid down in the preceding section:
\begin{enumerate}
    \item All the observables are publicly announced.
    \item Let $|\Psi_M\rangle=\frac{1}{\sqrt{2}}(|0\rangle+i  |2^N-1\rangle)$ be the reference state.  Bob$_1$  randomly prepares a state $|\Psi'_M\rangle$, obtained by a random measurement of one of the observables $\mathbb{O}$ from the set ${\cal S}_1$ on $|\Psi_M\rangle$ with an equal probability, followed by a masking transformation $\mathbb{U}_1$. That is,  $|\Psi'_M\rangle = \mathbb{U}_1\mathbb{\Pi}_{\mathbb{O}}^o|\Psi_M\rangle$, $\mathbb{O} \in {\cal S}_1$.  
  Then, he sends the state $|\Psi'_M\rangle$ to Bob$_2$.
    \item Bob$_2$ randomly  measures any one of the observables from the set ${\cal S}_2$ and performs a masking transformation $\mathbb{U}_{12}$ (which can be generated using the set of observables of first two Bobs) on the post-measurement state.
        Thereafter, he sends the transformed state to Bob$_3$.
    \item This process will continue till Bob$_N$ performs his measurement. This concludes the first round.
    \item This procedure is repeated for many rounds.
    \item After many such rounds are completed, the choices of observables are revealed for each round.
    \item The outcomes of such rounds, in which all the $N$-Bobs choose their respective observables from the set $\{\mathbb{Z}_1,\mathbb{Z}_2,\cdots,\mathbb{Z}_N\}$, are not revealed.
    \item The outcomes of rounds, in which all the $N$-Bobs choose their respective observable from the set $\{(\mathbb{X}_1,\mathbb{Y}_1),\cdots,(\mathbb{X}_N,\mathbb{Y}_N)\}$, are revealed. This data is used to check violation of Mermin's contextuality inequality given by (\ref{eq:5}).
    \item The outcomes of other rounds are discarded.
    \item If inequality (\ref{eq:5}) is violated, it implies the absence of eavesdropping and the sets of outcomes of $\mathbb{Z}_k $ work as a shared secure key.
    \item If inequality (\ref{eq:5}) is not violated, the presence of eavesdropping is indicated.
\end{enumerate}
A pictorial representation of procedure outlined is shown in figure (\ref{fig:mesh1}).

\subsection{Key generation rate}
\label{keyRate}
The key consists of outcomes of only those rounds  in which all the $N$-Bobs choose their respective observables from the set $\{\mathbb{Z}_1,\mathbb{Z}_2,\cdots,\mathbb{Z}_N\} $. Since Bob$_k$ randomly chooses one of the observables from the set $\{\mathbb{X}_k, \mathbb{Y}_k, \mathbb{Z}_k\}$, the probability for the event which generates the key is  $\dfrac{1}{3^N}$. The Shannon information for the above protocol is 1 bit. So, the average key generation rate is $\dfrac{1}{3^N}$ bit. We stress that if one takes into account only the sifted data as is usually done, the key generation rate is 1 bit.
 \section{ Can the location of Eve be pinpointed?}
 \label{Location}
   Violation of Mermin's contextuality inequality by all the $N$-Bobs assures the secrecy of the shared key, contingent on performing masking transformations. Then, a question can be asked: what happens if the Mermin's contextuality inequality is violated only by the $(N-M)$ Bobs? In such a case, can a key be shared securely among $(N-M)$ Bobs? We show that the answer to this question is in the negative.
  
     Let there be $N$ parties, {\it viz.}, Bob$_1$, Bob$_2$, $\cdots$, Bob$_{N}$. Bob$_k$ measures one of the observables  from the set $\{\mathbb{X}_k, \mathbb{Y}_k, \mathbb{Z}_k;1\le k \le N\}.$ However, Bob$_1$ enjoys the special status as he directly prepares the state which he might have obtained after performing measurement of the observable on the reference state $\ket{\Psi_M}$.  After $p$ rounds, there are $p$ strings of $N$ numbers that get generated as the outcomes of observables of $N$ Bobs. These strings are inevitably generated irrespective of eavesdropping. Employing these strings, the expectation values of the following observables can be calculated,
\begin{eqnarray}
    &\langle \mathbb{X}_k\rangle, \langle \mathbb{Y}_k\rangle, \langle \mathbb{Z}_k\rangle, \langle \mathbb{X}_k\mathbb{Y}_l\rangle, \langle \mathbb{X}_k\mathbb{X}_l\rangle,\cdots,\langle \mathbb{X}_1\cdots \mathbb{X}_{N}\rangle,\cdots.
\end{eqnarray}
With these expectation values, one can check whether Mermin's contextuality inequality gets violated for $N$-Bobs or for some $(N-M)$ Bobs.
  If the inequality is violated only by $(N-M)$ Bobs, it might appear that the secret key may still be shared among those $(N-M)$ Bobs.

However, consider a scenario, in which Eve is present between Bob$_M$ and Bob$_{M+1}$ and performs a measurement. It is due to the presence of Eve that Mermin's contextuality inequality is not violated among $(N-M-1)$ Bobs. Thus, Eve can obtain full information about the key. This implies key will be secure only if all the $N$-Bobs are violating Mermin's contextuality inequality. Thus, this analysis provides us with the information about the location of Eve. Although, we have considered Mermin's contextuality inequality explicitly, the similar analysis holds for any CQCP.
\section{QCP based on CHSH contextuality inequality}\label{Bell_Protocol} 
We have seen that in the Mermin's contextuality-based QCP, a single $2^N$ dimensional state is sent from Bob$_1$ to Bob$_N$. Although a multi-party nonlocal state is not required, the experimental limitations on generating higher dimensional states put constraints on the number of Bobs among which a key can be shared. Therefore, we ask a question: can this constraint be removed? The answer to the question is in the affirmative and can be obtained from the second class of QCPs, mentioned in section (\ref{Introduction}). However, it would require that every Bob can prepare a four-dimensional state.

We illustrate the new class by explicitly proposing QCP based on CHSH contextuality inequality. The QCP involves only a four-dimensional state, {\it irrespective of the number of Bobs}.  
  CHSH contextuality inequality \cite{Guhne10} for a single qudit is given as:
  \begin{equation}
  \label{Bellcontextualityineq}
   \big|\big\langle   \mathbb{A}_1(\mathbb{A}_2+\mathbb{A}'_2)+\mathbb{A}'_1(\mathbb{A}_2-\mathbb{A}'_2)\big\rangle\big|\leq 2,
  \end{equation}
  where $\mathbb{A}_1, \mathbb{A}'_1$ and $\mathbb{A}_2, \mathbb{A}'_{2}$ are observables acting on the space of a single qudit system. For the special case of a four-level system, we make the specific choice of observables as follows,
  \begin{eqnarray}
  \mathbb{A}_1=\mathbb{X}_1 ,~ \mathbb{A}'_1 = \mathbb{Z}_1 ;~ \mathbb{A}_2= \frac{\mathbb{X}_2+\mathbb{Z}_2}{\sqrt{2}},~\mathbb{A}'_2 = \frac{\mathbb{X}_2-\mathbb{Z}_2}{\sqrt{2}}.
  \end{eqnarray}
   The corresponding contextuality inequality (\ref{Bellcontextualityineq}) takes the form,
   \begin{equation}\label{CHSH}
      \big|\big\langle\mathbb{X}_1\mathbb{X}_2+\mathbb{Z}_1\mathbb{Z}_2\big\rangle\big|\leq \sqrt{2},
  \end{equation}
    and gets maximally violated by the state $\ket{\Psi_B}=\frac{1}{\sqrt{2}}(|1\rangle-|2\rangle)$. 
    
\begin{figure}[ht!]
    \centering
    \includegraphics[width=0.8\textwidth]{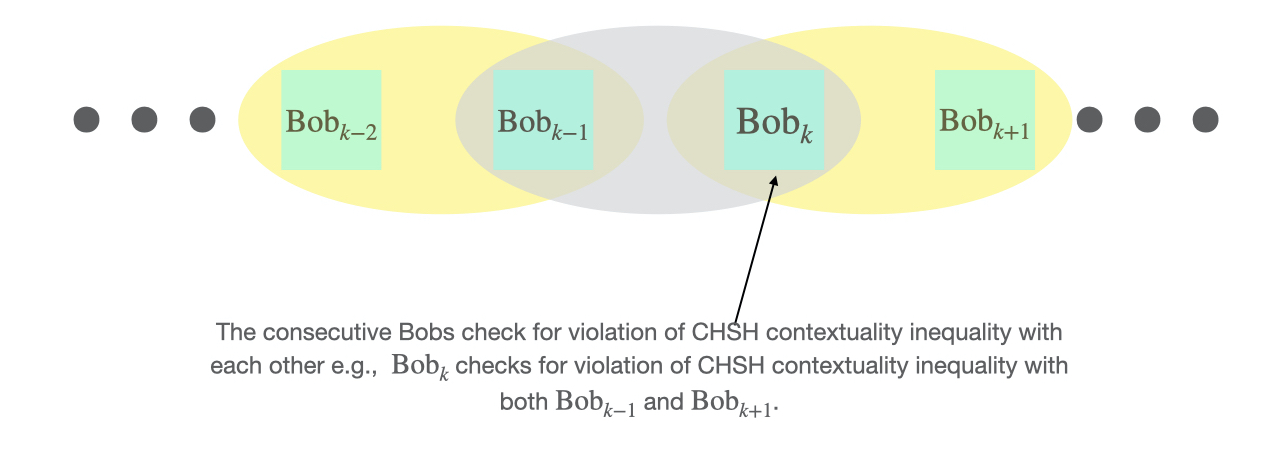}
    \caption {Grouping of consecutive parties in QCP based on CHSH contextuality inequality. }
    \label{fig:Bell2}
\end{figure}

    In this protocol, two consecutive parties are grouped, as shown in figure (\ref{fig:Bell2}). The odd numbered parties (i.e., Bob$_1$, Bob$_3$, $\cdots$) and the even numbered parties (i.e., Bob$_2$, Bob$_4$, $\cdots$) choose their observables from the sets, 
    \begin{eqnarray}
    {\cal S}_o & \equiv \Big\{\mathbb{X}_1, \frac{\mathbb{X}_1+\mathbb{Z}_1}{\sqrt{2}},\mathbb{Z}_1\Big\},~{\rm and},~
    {\cal S}_e  \equiv \Big\{\frac{\mathbb{X}_2+\mathbb{Z}_2}{\sqrt{2}}, \mathbb{Z}_{2}, \frac{-\mathbb{X}_2+\mathbb{Z}_{2}}{\sqrt{2}}\Big\},
\end{eqnarray}
 respectively.
The QCP based on CHSH contextuality inequality is as follows:
\begin{enumerate}
\item Let $\ket{\Psi_B}=\frac{1}{\sqrt{2}}(|1\rangle-|2\rangle)$ be the reference state. Bob$_1$ randomly prepares a state $|\Psi_1\rangle$, obtained by a random measurement of one of the observables $\mathbb{O}_1$ from the set ${\cal S}_o$ on $|\Psi_B\rangle$ with an equal probability, followed by a masking transformation $\mathbb{U}_1$. That is, 
$|\Psi_1\rangle = \mathbb{U}_1\mathbb{\Pi}_{{\mathbb O}_1}^{i}|\Psi_B\rangle$,

  Then, he sends the state $|\Psi_1\rangle$ to Bob$_2$.
\item $\rm{Bob}_2$ randomly measures one of the observables $\mathbb{O}_2$ from the set ${\cal S}_e$ on the state $|\Psi_1\rangle$, and obtains an outcome, say,  $j$. He should not send the post-measurement state to Bob$_3$ due to the masking transformation performed by Bob$_1$, which alters the correlations between observables of the sets $\mathcal{S}_o$ and $\mathcal{S}_e$.
\item   $\rm{Bob}_2$ prepares the state, $\ket{\Psi_2}= \mathbb{U}_2{\mathbb\Pi}_{{\mathbb O}_2}^j\ket{\Psi_B}$, and sends it to Bob$_3$.  Here, $\mathbb{U}_2$ is a masking transformation.
\item $\rm{Bob}_3$ performs measurement of one of the observables $\mathbb{O}_3$ from the set $\mathcal{S}_o$ on $|\Psi_2\rangle$, and obtains an outcome, say, $l$.
\item  $\rm{Bob}_3$ prepares the state, $\ket{\Psi_3}= \mathbb{U}_1'{\mathbb\Pi}_{{\mathbb O}_3}^l\ket{\Psi_B}$, and sends it to Bob$_4$. Here, $\mathbb{U}_1'$ is a masking transformation.
\item This process goes on till Bob$_{N}$ performs his measurement. After the measurement of Bob$_N$, one round is completed. The same process is repeated for many rounds. 
\item After that, the choice of observables for each round is made public. The outcomes of those rounds, in which all the $N$-Bobs have chosen their observables from either of the sets $\{\frac{\mathbb{X}_1+\mathbb{Z}_1}{\sqrt{2}},\frac{\mathbb{X}_2+\mathbb{Z}_2}{\sqrt{2}}\},\{\mathbb{Z}_1,\mathbb{Z}_2\}$ are not revealed. The outcomes for all other rounds are revealed. 
\item Consecutive Bobs check for violation of CHSH contextuality inequality (\ref{CHSH}) with each other, i.e., Bob$_k$ checks for violation of (\ref{CHSH}) with Bob$_{k-1}$ and Bob$_{k+1}$ as shown in figure (\ref{fig:Bell2}).
\item If the CHSH contextuality inequality gets violated between all the pairs, the outcomes of those runs in which all Bobs have chosen their observables from either of the sets $\{\frac{\mathbb{X}_1+\mathbb{Z}_1}{\sqrt{2}},\frac{\mathbb{X}_2+\mathbb{Z}_2}{\sqrt{2}}\},\{\mathbb{Z}_1,\mathbb{Z}_2\}$,  act as a shared secret key.
\end{enumerate}

\begin{figure}[ht!]
    \centering
    \includegraphics[width=0.8\textwidth]{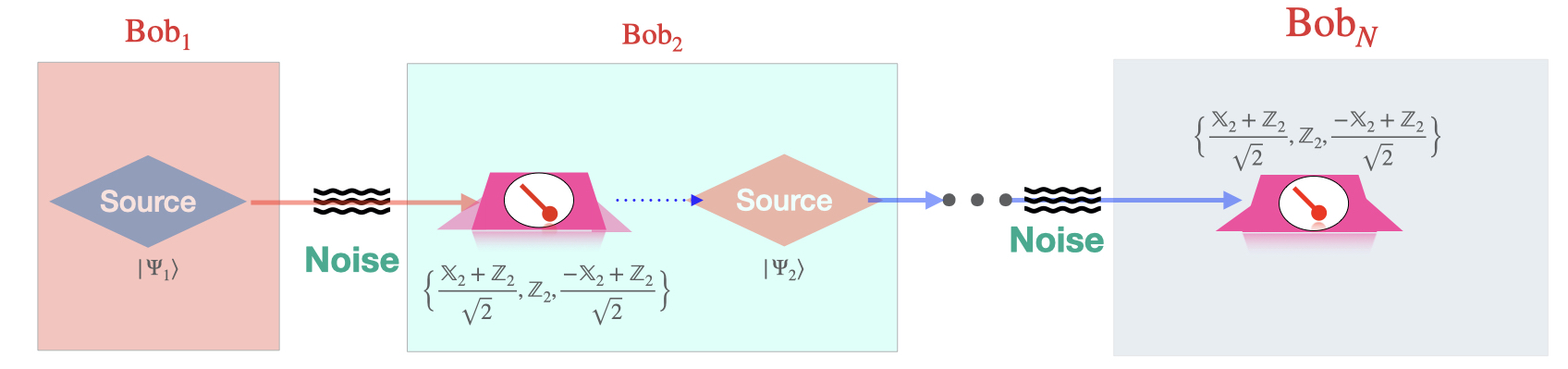}
    \caption{Schematic representation of QCP based on CHSH contextuality inequality. 
       }
    \label{fig:Bell1}
\end{figure}

This concludes the description of QCP based on CHSH contextuality inequality. 
A pictorial representation of CHSH contextuality-based QCP is given in figure (\ref{fig:Bell1}).


\subsection{Key generation rate}
The key generation rate of this protocol is twice that of  the Mermin's contextuality based QCP. The higher key generation rate owes to the fact that the outcomes of two observables per Bob contribute to the key generation whereas, in Mermin-based QCP, the outcome of only one observable per Bob contributes to it.  The probability that all the $N$-Bobs choose the same observables is $\frac{1}{3^N}$. Since, measurement outcomes of the  two sets of observables, i.e., $\{\frac{\mathbb{X}_1+\mathbb{Z}_1}{\sqrt{2}}, \frac{\mathbb{X}_2+\mathbb{Z}_2}{\sqrt{2}} \};~\{\mathbb{Z}_1, \mathbb{Z}_2\}$ are fully correlated, the key generation rate is  $\frac{2}{3^N}$.

 
  \section{Outlook for implementation using OAM states of light}
  \label{OAM}
 In this section, we examine the feasibility of experimental implementation of {the QCPs proposed in sections (\ref{Protocol}) and (\ref{Bell_Protocol})} using OAM states, although it may not be realised in the immediate future. Depending on the symmetry of laser resonant cavity , Laguerre-Gauss modes \cite{Allen}, Bessel modes\cite{andrews} etc., which carry the OAM states of light are experimentally generated. Laguerre-Gauss modes, which are the solutions of paraxial Helmholtz wave equation, have attracted a lot of experimental attention\cite{mode,mirhosseini,vaziri} ever since they were theoretically proposed \cite{Allen}. These modes are represented as $LG_{p,l}$, with $p$ being the radial index and $l$, the azimuthal index. The implementation of the protocols employing OAM states requires only linear optics, which makes it realizable with coherent pulses as well. As we are only interested in OAM degree of freedom, we freeze the radial index to be equal to zero, and write the corresponding states as $|LG_l\rangle$.
 
 Experimental implementation of the proposed QCPs involve three steps, (i) preparation of the state , (ii) measurement of observables, and (iii) unitary transformations performed on post-measurement state. We outline the implementation of these steps sequentially as follows.
   
\noindent \textbf{Preparation of state:}
In Mermin's contextuality based QCP, we require an initial state,  $|\Psi_M'\rangle$  belonging to a $2^N$ dimensional Hilbert space.  A state with a definite value of $`l'$ or their superposition can be prepared using a spatial light modulator (SLM), which is basically an automated liquid crystal device.
So, Mermin-based CQCP can be implemented with single qudit states (e.g., OAM states) whose generation and manipulation are experimentally easier as compared to the multi-party nonlocal states. For example, OAM states with $l_{\rm{max}} = 100$ have been prepared with $90\%$ purity \cite{Sroor20}, while a four-photon GHZ states have been  prepared with $81\%$ purity and $87\%$ fidelity \cite{4photonGHZ}.  With a $100$ dimensional state, Mermin's  CQCP with only six Bobs can be implemented. However, this limitation no longer exists in the QCP based on CHSH contextuality inequality, which requires only a four-dimensional state irrespective of the number of Bobs involved.\\
\textbf{Measurement of observables:} 
The observables in our protocols are dichotomic. Thus, they apportion the Hilbert space into two orthogonal spaces of equal dimensions. Again, SLM can be employed for projecting a state on these eigenprojections.

\noindent\textbf{Unitary transformations:}  These transformations on OAM states can be performed using beam splitters and Dove prisms\cite{leach2004interferometric}, or more generically through the programmable holographic techniques \cite{Wang17}.  This method allows, in principle, the implementation of any unitary transformation on the OAM modes.
 
 Of course, a complete analysis of real-life implementation of these QCPs would also require incorporation of various other effects such as non-ideal behaviour of experimental equipment and turbulence in atmosphere, which has not been considered in this paper.

Finally we note that irrespective of physical system employed for implementation, though we employ $ d^N$ dimensional quantum states and measurement operators, the key consists of only $d$ symbols.

   \section{Error analysis}
   \label{Error}
   Errors may arise in communication systems due to (i) noisy channels, (ii) imperfections in the preparation of states, (iii) noisy operations, or, (iv) through noisy measurements.
  In this section, we study the vulnerability of the proposed protocols for imperfect preparation of the state and noisy measurements. We take explicit examples of Mermin and CHSH contextuality-based QCPs. Effect of these errors on the key generation rate is examined. Although masking transformations are required at each step in the protocols, but to make discussion easier, we ignore them for this analysis. We also consider noise only in those states and detectors which are used in the process of key generation, as we are only interested in the effect of these errors on the key generation rate.
  
Naturally,   the asymptotic key generation rate of the protocol is given by the minimum of mutual information when all parties are considered pairwise, i.e.,
  \begin{equation}
 \label{eq:keyrate}
   r=\min_{\substack{1\leq i,j\leq N\\ i < j}}[I(B_i,B_j)],
  \end{equation}
 where $I(B_i, B_j)$ is the mutual information between Bob$_i$ and Bob$_j$.
 
  Please note that the equation (\ref{eq:keyrate}) falls short of yielding the secret key rate. The effects of various attacks on two--party quantum key distribution protocols have been studied (see \cite{scarani2009security} and references therein). Extension of those results to the contextuality--based QCPs presented in this work will be taken up separately. The analysis, given in this section, has been restricted to observe the effect of noisy preparation and measurements on the key generation rate. 
  \subsection{Imperfect preparation of the state}
  \subsubsection{Noise in two-dimensional subspace}
  \textit
  {a)  Mermin's contextuality inequality based QCP :}\vspace{0.15cm}\\
  \label{Model1}
  Recall that the protocol starts with Bob$_1$ preparing his system in the state $\ket{\Psi_1}=\mathbb{\Pi}^o_{\mathbb{O}}|\Psi_M\rangle$, $\mathbb{O} \in {\cal S}_1$. The only measurement that contributes to key generation is that of $\mathbb{Z}_1$. Bob$_1$ wants to prepare  either of the states $\ket{0}$ or $\ket{D-1}$ for the respective outcomes $+1$ and $-1$ of the observable $\mathbb{Z}_1$. However, as the preparation of the state can be noisy, he would end up preparing either of the following states, given as,
  \begin{eqnarray}\label{eq:M1}
      \rho_{1M}=(1-\epsilon_1)\ket{0}\bra{0}+\epsilon_1\ket{D-1}\bra{D-1},\nonumber\\~
      \rho_{2M}=(1-\epsilon_2)\ket{D-1}\bra{D-1}+\epsilon_2\ket{0}\bra{0},
  \end{eqnarray}
  where $\epsilon_1,~\epsilon_2$ are small.
   The effect of this imperfection in the preparation of the state is equivalent to the binary flip channel employed classically, where $\ket{0}$ would flip to $\ket{D-1}$ with probability $\epsilon_1$ and $\ket{D-1}$ to $\ket{0}$ with probability $\epsilon_2$.
  
  In this QCP, only Bob$_1$ prepares a state. So, the effect of noise will be prominent in the channel between Bob$_1$ and Bob$_2$. The correlation among rest of Bobs is perfect. Therefore, the mutual information between Bob$_1$ and Bob$_2$ will be minimum and hence determines the key generation rate, $r_M$, i.e.,
  \begin{equation}
      r_M = I(B_1, B_2).
  \end{equation}
  
\begin{figure}
  \centering
  \includegraphics[width=.6\linewidth]{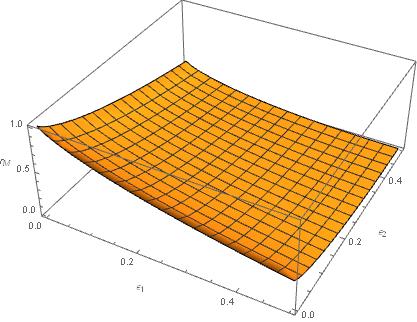}
  \caption{ Variation of the key generation rate $r_M$ w.r.t noise in the state ($\epsilon_1,~\epsilon_2$) for Mermin's contextuality based QCP for $N=3$. For $\epsilon_1,~\epsilon_2=0$, the key generation rate attains its ideal value, i.e., the one in the noiseless scenario.}
   \label{fig:Model1}
\end{figure}

  As an illustration, we consider the case of three Bobs and explicitly plot the key rate $(r_M)$ with respect to $(\epsilon_1,\epsilon_2)$ in figure (\ref{fig:Model1}).

   A similar analysis for QCP based on CHSH contextuality inequality is as follows.\\
   \textit{b) CHSH contextuality inequality based QCP :}\vspace{0.15cm}\\
     In this QCP, a key is generated by the measurements of the two sets of observables, $\big(\big\{\frac{\mathbb{X}_1+\mathbb{Z}_1}{\sqrt{2}},~\frac{\mathbb{X}_2+\mathbb{Z}_2}{\sqrt{2}}\big\};\{\mathbb{Z}_1,\mathbb{Z}_2\}\big)$, as explained in section (\ref{Bell_Protocol}). We analyse one set of observables $\{\mathbb{Z}_1,\mathbb{Z}_2\}$. A similar calculation can be done for the other set of observables as well.

  Bob$_1$ prepares either of the states $\ket{1}$ or $\ket{2}$ for the respective outcomes $+1, -1$ of the  observable $\mathbb{Z}_1$.
  As the preparation of the states is noisy, he would end up preparing the following states: 
  \begin{eqnarray}
      \rho_{1B}=(1-\epsilon_1)\ket{1}\bra{1}+\epsilon_1\ket{2}\bra{2},\nonumber\\
      \rho_{2B}=(1-\epsilon_2)\ket{2}\bra{2}+\epsilon_2\ket{1}\bra{1},
  \end{eqnarray}
which again admits an analysis via binary flip channel. Recall that in this QCP, each Bob has to prepare the state.

   We consider the case of three Bobs.
By using equation (\ref{eq:keyrate}), the key generation rate is given by the minimum of mutual information between any two Bobs, which in this case is $I(B_1,B_3)$, i.e.,
   \begin{equation}
     r_B= I(B_1,B_3).
   \end{equation}
  
\begin{figure}
  \centering
  \includegraphics[width=.6\linewidth]{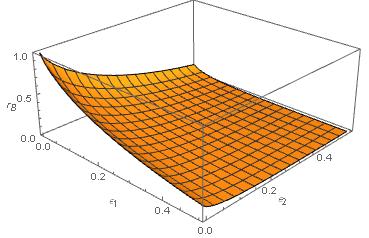}
\caption{ Variation of the key generation rates $r_B$ w.r.t noise in the state ($\epsilon_1,~\epsilon_2$) for CHSH contextuality based QCP in respectively  for $N=3$. For $\epsilon_1,~\epsilon_2=0$, the key generation rate attains its ideal value, i.e., the one in the noiseless scenario.}
  \label{fig:test}
\end{figure}

  The effect of noisy preparation of states for CQCPs based on both Mermin and CHSH contextuality inequalities can be seen in figures (\ref{fig:Model1}) and (\ref{fig:test}) respectively. The key generation rate falls more rapidly for CQCP based on CHSH inequality. This is a signature of  greater noise since both the Bobs ({\it viz.}, Bob$_1$ and Bob$_2$) prepare the state afresh with a noisy apparatus.
   

  \subsubsection{White noise}
\begin{figure}
  \centering
  \includegraphics[width=.5\linewidth]{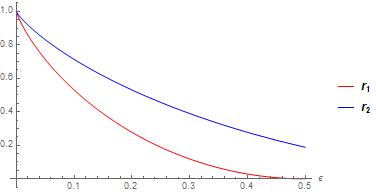}
\caption{Variation of key generation rates $r_1=r_M~\rm{at}~\epsilon_1,\epsilon_2=\epsilon$, and $r_2$ w.r.t noise ($\epsilon$) in the state for QCP based on Mermin contextuality inequality.}
\label{fig:white}
\end{figure}

In this section, we consider a scenario in which the  state gets contaminated by the white noise. That is, when Bob$_1$ wants to prepare the states $\ket{0}\bra{0}$ or $\ket{D-1}\bra{D-1}$, he would instead end up preparing states, 
\begin{eqnarray}
  &  \rho'_1&=(1-\epsilon)|0\rangle\langle 0| +\frac{\epsilon}{8}\mathbb{1},\nonumber\\
      &  \rho'_2 &=(1-\epsilon)|D-1\rangle\langle D-1| +\frac{\epsilon}{8}\mathbb{1},
\end{eqnarray}
respectively. For this kind of noisy preparation, the key generation rate, $r_2$, is found to be:
\begin{equation}
   r_2=I(B_1,B_2),
\end{equation} 
For the case of three Bobs, the key generation rate is plotted against $\epsilon$ in figure (\ref{fig:white}). For the sake of comparison, we have also plotted $r_1=r_M\vert_{\epsilon_1,\epsilon_2=\epsilon}$ in the same plot.

A similar analysis can be done for QCP based on CHSH contextuality inequality.
  \subsection{Imperfect detectors}
  We consider an example of imperfect detectors, that register correctly with a probability $(1-\eta)$ and incorrectly (i.e., in the orthogonal subspace), with a probability $\eta$. Such a detector gets represented by the operator,
  \begin{eqnarray}\label{eq:det}
            &&\mathbb{M}_{\mathbb{Z}_k}^{+}=(1-\eta)\mathbb{\Pi}_{\mathbb{Z}_k}^++\eta\mathbb{\Pi}_{\mathbb{Z}_k}^-,\nonumber\\
     &&\mathbb{M}_{\mathbb{Z}_k}^-= \eta\mathbb{\Pi}_{\mathbb{Z}_k}^+ +(1-\eta)\mathbb{\Pi}_{\mathbb{Z}_k}^-;~k \in \{2, \cdots, N\}.
     \end{eqnarray}

\begin{figure}
\centering
  \includegraphics[width=.6\linewidth]{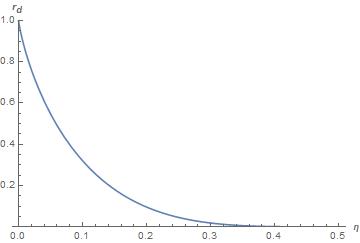}
  \caption{Variation of key generation rate $r_d$ w.r.t noise in detector $\eta$.}
   \label{fig:D1}
\end{figure}

 The effect of noise in equation (\ref{eq:det})  can be modelled by a binary symmetric channel, endowed with a probability of flip, $\eta.$ In the protocol for three Bobs, since Bob$_2$ and Bob$_3$ both perform measurements, the noise can be modelled by two cascaded binary symmetric channels. The key generation rate, $r_d$, is found to be,
\begin{equation}
     r_d=I(B_2,B_3),
  \end{equation}
 which is plotted in figure (\ref{fig:D1}).

 A similar analysis can be done for CHSH contextuality based QCP.
 
  \subsection{Both detectors and states noisy}
  \subsubsection{Model I}
  \label{M1}
\begin{figure}
  \includegraphics[width=.5\linewidth]{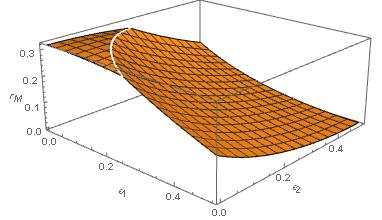}
  \caption{Variation of key generation rate ($r_M$) for a particular $\eta=0.1$ (noise in the detector) w.r.t noise in the preparation of state, ($\epsilon_1,~\epsilon_2$),  for QCP based on Mermin contextuality inequality.}
   \label{fig:11.3.1}
\end{figure}

  \textit{ Mermin's contextuality based QCP}\vspace{0.15cm}\\
  Consider a case in which both the state and the measurements are noisy. The noisy measurement is such that it misreads the signal for one projection operator as the other. That is to say, for some fraction of times, it reads $\mathbb{\Pi}^+$ as $\mathbb{\Pi}^-$ and vice versa. The noisy state sent by Bob$_1$ to Bob$_2$ for $+1$ or $-1$ outcome of $\mathbb{Z}_1$ respectively are 
  \begin{eqnarray}
     \rho_{1M} =  (1-\epsilon_1)|0\rangle\bra{0}+\epsilon_1|D-1\rangle\bra{D-1},\nonumber\\
     \rho_{2M} =  (1-\epsilon_2)|D-1\rangle\bra{D-1}+\epsilon_2|0\rangle\bra{0}.\nonumber
   \end{eqnarray}
  The noisy measurement operators are:
  \begin{eqnarray}            &&\mathbb{M}_{\mathbb{Z}_k}^{+}=(1-\eta)\mathbb{\Pi}_{\mathbb{Z}_k}^++\eta\mathbb{\Pi}_{\mathbb{Z}_k}^-,\nonumber\\
     &&\mathbb{M}_{\mathbb{Z}_k}^-= \eta\mathbb{\Pi}_{\mathbb{Z}_k}^+ +(1-\eta)\mathbb{\Pi}_{\mathbb{Z}_k}^-;~k \in \{2, \cdots, N\}.
     \end{eqnarray}
             Consider a special case of three Bobs\footnote{For higher number of parties, calculations are not difficult, but become tedious. Hence, the choice of three Bobs.}. The key generation rate is given by the minimum of pairwise mutual information. For this model, depending upon the amount of noise, we observe the interplay between mutual information between Bob$_1$ and Bob$_2$, $I(B_1,B_2)$ and mutual information between Bob$_2$ and Bob$_3$, $I(B_2,B_3)$. The key generation rate $r_M$ is found to be
             \begin{equation}
       r_M= \min[I(B_1,B_2), I(B_2,B_3)].
             \end{equation}
            We have plotted this key generation rate $ r_M$ for a fixed  $\eta=0.1$ (noise in the detector) w.r.t noise in the preparation of state, ($\epsilon_1,~\epsilon_2$), in figure (\ref{fig:11.3.1}). The left side of discontinuity in the figure represent the region over which $I(B_2,B_3)$ is minimum and on the right side  $I(B_1,B_2)$ is minimum.

       In a similar way, calculation for CHSH contextuality based QCP can be performed.
     \subsubsection{Model II}
     \textit{Mermin's contextuality based QCP }\\
      In this model, we consider an imperfect preparation of the state and lossy detectors.
      The detector misses a signal with a probability $(1-\eta)$ and is represented as: 
      \begin{equation}          \mathbb{M}_{\mathbb{Z}}^{\pm}=\eta\mathbb{\Pi}_\mathbb{Z}^{\pm}.
           \end{equation}
        Consider a particular case of three Bobs, for which Bob$_1$ prepares the two states with an  equal probability:
        \begin{eqnarray}
            \rho_1 = (1-\epsilon_1)|0\rangle\bra{0}+\epsilon_1 |7\rangle\bra{7},\nonumber\\
            \rho_2 = (1-\epsilon_2)|7\rangle\bra{7}+\epsilon_2|0\rangle\bra{0}.
        \end{eqnarray}

        The effect of this channel on the protocol can be modelled by binary flip channel and erasure channel. The minimum mutual information is found to be $I(B_1,B_3)$ and hence act as the key generation rate, $r_M$, i.e.,
        \begin{equation}
           r_M= I(B_1,B_3). 
        \end{equation}

  A similar analysis can be done for CHSH contextuality based QCP for which the key generation rate, $r_B$, is found to be:
  \begin{equation}
      r_B=I(B_1,B_3).
  \end{equation}
 
\begin{figure}
  \centering
  \includegraphics[width=.6\linewidth]{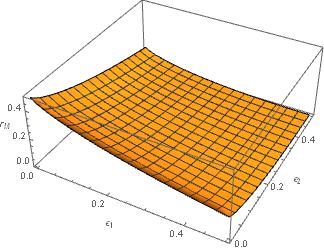}
  \caption{ Variation of the key generation rate $r_M$ w.r.t noise in the state ($\epsilon_1$ and $\epsilon_2$) for model II for $\eta =0.7$ noise in detector for QCP based on Mermin contextuality inequality.}
   \label{fig:Model4}
\end{figure}

\begin{figure}
  \centering
  \includegraphics[width=.6\linewidth]{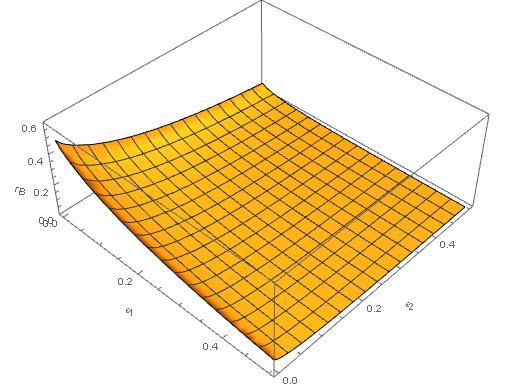}
\caption{ Variation of the key generation rate $r_B$ w.r.t noise in the state ($\epsilon_1$ and $\epsilon_2$) for model II for $\eta =0.7$ noise in detector for QCP based on CHSH contextuality inequality.
}
\label{fig:test1}
\end{figure}

 Effect of this noisy channel on the key generation rate is plotted in figures (\ref{fig:Model4}) and (\ref{fig:test1}) for QCP based on Mermin and CHSH contextuality inequality respectively.
      


\section{Conclusion}
\label{Conclusion}
In summary, we have laid down the procedure to obtain device-dependent CQCP from any NQCP. These protocols do not involve multiparty nonlocal states. Two classes of QCPs are proposed through the examples of Mermin’s and CHSH contextuality inequality. The latter is more beneficial as it alleviates the constraint on the number of Bobs who can participate in the key sharing process.

Finally, the equivalence of multi-party nonlocality and single qudit contextuality which we have shown here, can be used to develop contextuality-based quantum secure direct communication protocols from nonlocality-based quantum secure direct communication protocols.

\acknowledgement
 The authors would like to thank  the anonymous reviewers for their valuable comments that help enhancing the quality of the paper. Rajni thanks UGC for funding her research. Sooryansh  thanks the Council for Scientific and Industrial Research (Grant no. -09/086 (1278)/2017-EMR-I) for funding his research.

\section*{Author Contribution Statement}
All the authors contributed equally in all respects.
\begin{appendix}
     \section{Mermin's nonlocality inequality as contextuality inequality}
          \label{Mermin_contextuality}
        The interrelation between the two-party nonlocality and single-qudit contextuality,  shown in the section (\ref{Bell_contextuality}) for CHSH inequality, continues to hold even for multi-party nonlocality inequalities. We illustrate it through the example of Mermin's inequality. For that, we briefly recapitulate  Mermin's inequality for an $N$-party system. 
        
        Mermin's inequality distinguishes the states admitting completely factorisable local hidden variable model from nonlocal states. 
          
          Consider a pair of dichotomic observables $\{A_{j}, A'_{j} \}$ in the space of the $j^{\rm th}$ party ($ j\in\{1, \cdots, N\}$). The inequality is as follows \cite{Mermin90}:
          \begin{eqnarray}\label{Mermin}
    {\cal M}_N && =\frac{1}{2i}\Bigg\vert\bigg\langle\prod_{j=1}^{N} (A_j+i A'_j)-\prod_{j=1}^{N}(A_j-i A'_j)\bigg\rangle\Bigg\vert\leq c\nonumber\\
 c&&= 2^{N/2},\quad N={\rm even},\nonumber\\
 &&= 2^{\frac{N-1}{2}},\quad N= {\rm odd}; N \geq 3.
\end{eqnarray}
 The observables corresponding to different parties commute and $[A_k,A'_k] \neq 0$.

For the special case of an $N$ qubit system, the inequality (\ref{Mermin}) gets maximally violated by the GHZ state,
\begin{eqnarray}\label{GHZ_N}
    |\phi_N\rangle=\dfrac{1}{\sqrt{2}}\bigg(|0\rangle ^{\otimes N}+i|1\rangle ^{\otimes N}\bigg),
\end{eqnarray}
for the following choice of the observables:
\begin{eqnarray}
    A_k =X_k;~A'_k=Y_k.
\end{eqnarray}
It is pertinent to illustrate the contextual behaviour of $|\phi_N\rangle$ to substantiate our claim. For that purpose, we consider the following two contexts:\\
 \textbf{Context 1 ($Z_1 Z_2\cdots Z_{N-1}Z_N $):} Suppose the outcome of $Z_1$ is $+1(-1)$. Then, the outcomes of observables $Z_2, Z_3, \cdots, Z_N$ will be $+1(-1)$ with unit probability. \\
  \textbf{Context 2 ($Y_1Z_2\cdots Z_N$):} Let the outcome of the measurement of $Y_1$ be once again, +1. Following it, the measurement of observable $Z_2$ will yield $+1$ or $-1$ with equal probability. Thereafter, the measurement of $Z_3, \cdots, Z_N$ will definitely yield +1. Thus, the outcome of $Z_2$ depends on the set of commuting observables it is measured with, i.e., it depends on the {\it context}. \\
  Exactly in the same manner as in the section (\ref{Bell_contextuality}), the observables and the state in the $2^N$ dimensional Hilbert space can be identified. 
\begin{equation}
  |\phi_N\rangle \mapsto |\Phi\rangle=\dfrac{1}{\sqrt{2}}(|0\rangle +i |2^N-1\rangle)
\end{equation}
The equivalent observables $\{\mathbb{Z}_1,\cdots, \mathbb{Z}_N,\mathbb{Y}_1 \}$ can be obtained by following the prescription in point (2) and (3) of section (\ref{Notation}), and they satisfy the same commutation relations as $\{{Z}_1,\cdots, {Z}_N, {Y}_1 \}$ because of operator isomorphism.\\
Thus, Mermin's inequality probes contextuality in a qudit of appropriate higher dimension and can be referred to as Mermin's contextuality inequality.

\section{Illustration of eavesdropping}
\label{Illustration}
To illustrate the point that how Eve obtains information about the key without disturbing the violation of contextuality inequality, we consider the QCP based on Mermin's contextuality inequality.
   Consider an example of three parties, Bob$_1$, Bob$_2$ and Bob$_3$. We choose following set of observables:
  \begin{eqnarray}
     &&\mathbb{Z}_1=\ket{0}\bra{0}+\ket{1}\bra{1}+\ket{2}\bra{2}+\ket{3}\bra{3}\nonumber\\
      &&~~~~~-\Big(\ket{4}\bra{4}+\ket{5}\bra{5}+\ket{6}\bra{6}+\ket{7}\bra{7}\Big)\nonumber\\
     &&\mathbb{Z}_2=\ket{0}\bra{0}+\ket{1}\bra{1}+\ket{4}\bra{4}+\ket{5}\bra{5}\nonumber\\
      &&~~~~~-\Big(\ket{2}\bra{2}+\ket{3}\bra{3}+\ket{6}\bra{6}+\ket{7}\bra{7}\Big)\nonumber\\
     &&\mathbb{Z}_3=\ket{0}\bra{0}+\ket{2}\bra{2}+\ket{4}\bra{4}+\ket{6}\bra{6}\nonumber\\
      &&~~~~~~-\Big(\ket{1}\bra{1}+\ket{3}\bra{3}+\ket{5}\bra{5}+\ket{7}\bra{7}\Big)\nonumber\\
    &&\mathbb{X}_1=\ket{0^+}\bra{0^+}+\ket{1^+}\bra{1^+}+\ket{2^+}\bra{2^+}+\ket{3^+}\bra{3^+}\nonumber\\
      &&~~~~~-\Big(\ket{0^-}\bra{0^-}+\ket{1^-}\bra{1^-}+\ket{2^-}\bra{2^-}+\ket{3^-}\bra{3^-}\Big),
     \end{eqnarray}
     where the symbols $\ket{l^{\pm}}$ are defined as,
      \begin{eqnarray}
     \ket{l^\pm}=\frac{1}{\sqrt{2}}(\ket{l}\pm\ket{l+4});~~~ l \in \{0, 1, 2, 3\}
      \end{eqnarray}
 Consider the following two situations, one in which there is no eavesdropping, the other in which Eve is present.
  \begin{enumerate}
     \item \textbf{Case I:} Following the steps of Mermin's contextuality based QCP,  let Bob$_1$ prepares a state $\ket{\Psi_M'}=\ket{0}$ for the +1 outcome of $\mathbb{Z}_1$ on the reference state $\ket{\Psi_M}$. He sends this state $\ket{\Psi_M'}$ to Bob$_2$. If Bob$_2$ measures, say, $\mathbb{Z}_2$ on this state, he obtains the outcome +1 with unit probability. Thereafter, he sends the post-measurement state to Bob$_3$. Bob$_3$ measures, say, $\mathbb{Z}_3$, he is bound to get outcome +1 with unit probability.
    \item  \textbf{Case II:} Let there be an eavesdropping between Bob$_2$ and Bob$_3$. Eve measures an observable, $\mathbb{X_1}$. She obtains $\pm 1$ with equal probability and the state collapses to $\ket{\Phi^{\pm}}=\frac{1}{\sqrt{2}}(\ket{0}\pm\ket{4})$. She sends the post-measurement state to Bob$_3$. Bob$_3$ measures an observable $\mathbb{Z}_3$, as in the previous case, and obtains outcome +1 with unit probability.\\
 Thus, the measurement of Eve does not affect the outcome of Bob$_3$'s measurement. Due to this, Bob$_3$ will never detect presence of Eve. In this way, Eve can obtain full information about the key by choosing appropriate observable without being detected.
  \end{enumerate}

\end{appendix}
\bibliographystyle{unsrt}

\end{document}